\documentclass
[superscriptaddress,secnumarabic,amssymb,amsmath,nobibnotes,aps,prd,showkeys,showpacs,nofootinbib,twocolumn]{revtex4}%
\usepackage{graphicx}
\usepackage{epsf}
\usepackage{bm}
\usepackage{amsmath}
\usepackage{amsfonts}
\usepackage{amssymb}%
\providecommand{\U}[1]{\protect\rule{.1in}{.1in}}

\newcommand{\be}{\begin{equation}}
\newcommand{\ee}{\end{equation}}

\newcommand{\mincir}{\raise
-3.truept\hbox{\rlap{\hbox{$\sim$}}\raise4.truept\hbox{$<$}\ }}
\newcommand{\magcir}{\raise
-3.truept\hbox{\rlap{\hbox{$\sim$}}\raise4.truept\hbox{$>$}\ }}

\begin{document}
\title{Dust fluid component from Lie symmetries in Scalar field Cosmology}
\author{Andronikos Paliathanasis}
\email{anpaliat@phys.uoa.gr}
\affiliation{Instituto de Ciencias F\'{\i}sicas y Matem\'{a}ticas, Universidad Austral de
Chile, Valdivia 5090000, Chile}
\affiliation{Institute of Systems Science, Durban University of Technology, PO Box 1334,
Durban 4000, Republic of South Africa}

\begin{abstract}
We show that in scalar-field cosmology, a dust fluid  follows as quantum
corrections from solutions of the Wheeler-DeWitt equation generated by Lie
symmetries. The energy density of the dust fluid is related with the frequency
of the wavefunction.

\end{abstract}
\maketitle
\date{\today}

%

\bigskip

Lie groups constitute an important tool for the study of natural systems and
are applied in all areas of physics \cite{lie1,lie2,lie3}. In gravity group
invariant transformations play a significant role in every stage of the
theory, from the definition of the natural space until the existence of
solution for the field equations. The application of Lie groups in the field
equations is the subject in which we are interested for this work.
Specifically we use the Lie group invariants to find solutions of the
Wheeler-DeWitt equation (WdW) for a scalar-field cosmological model while in
the context of the semiclassical approach of Bohmian mechanics the quantum
potential is determined.

Consider the Action Integral for the gravitational field equations to be that
of a minimally coupled scalar-field cosmology without any other matter source,
that is,%
\begin{equation}
S=\int dx^{4}\sqrt{-g}\left[  R-\frac{1}{2}g^{\mu\nu}\phi_{;\mu}\phi_{;\nu
}+V(\phi)\right]  ,
\end{equation}
where for the underlying spacetime we assume that of spatially flat
Friedmann-Lema\^{\i}tre-Robertson-Walker spacetime with scale factor $a\left(
t\right)  $. Without loss of generality we consider the lapse function to be
$N\left(  t\right)  =1$.

From the moment that we selected the form of the line element which defines
the underlying geometry, we have assumed that the application of a
six-dimensional Lie algebra leaves invariant the parallel lines in the space.
Furthermore the zero spatial curvature tells us that the group is $E^{3}$,
that is, $\left\{  3A_{1}\otimes_{s}SO\left(  3\right)  \right\}  $, in
contrast to the $SO\left(  4\right)  $ of the space of constant curvature. The
isometry group passes to all the geometric quantities which are generated by
the metric tensor, that is, the action of the isometry group leaves invariant
the Einstein tensor. Furthermore we assume that the scalar field inherits as
symmetries the isometry group, which leads to $\phi=\phi\left(  t\right)  $
and describes the quintessence.

The gravitational field equations follow from the Euler-Lagrange equations of
the Lagrange function,%
\begin{equation}
L\left(  a,\dot{a},\phi,\dot{\phi}\right)  =-3a\dot{a}^{2}+\frac{1}{2}%
a^{3}\dot{\phi}^{2}-a^{3}V\left(  \phi\right)  ,
\end{equation}
and the constraint equation,%
\begin{equation}
\mathcal{H}\equiv-\frac{1}{12a}p_{a}^{2}+\frac{1}{2a^{3}}p_{\phi}^{2}%
+a^{3}V\left(  \phi\right)  =0, \label{lang.02}%
\end{equation}
which has been expressed in terms of the momenta~$p_{a}=-6a\dot{a}$
and$~p_{\phi}=a^{3}\dot{\phi}$.

The mathematical formula in the minisuperspace approach of the WdW equation
for our model of consideration is \cite{wdw1,wdw2}%
\begin{equation}
-\Delta\Psi+2a^{3}V\left(  \phi\right)  \Psi=0, \label{lang.01}%
\end{equation}
in which $\Delta=\frac{1}{\sqrt{\left\vert \gamma\right\vert }}\frac{\partial
}{\partial x^{i}}\left(  \sqrt{\left\vert \gamma\right\vert }\frac{\partial
}{\partial x^{j}}\right)  $ is the Laplacian operator of the minisuperspace
which is defined by the field equations.

It is important to mention here that, because the dimension of the
minisuperspace is two, equation (\ref{lang.02}) is invariant under conformal
transformations. That means that equation (\ref{lang.02}) follows also from
the quantization of the quantity $N\left(  \tau\right)  \mathcal{H}=0$, in
which $N\left(  \tau\right)  $ is any lapse function of the FLRW spacetime,
$dt=N\left(  \tau\right)  d\tau.$However, that is not true for a
minisuperspace of dimension greater than two in which the Laplacian $\Delta$
is replaced by the conformally invariant Laplace operator, $L_{\gamma}$, in
the definition of the WdW equation.

From the theory of differential equations we know that when there exists a Lie
group under which equation (\ref{lang.02}) is invariant closed-form solutions
of (\ref{lang.02}) exist. We assume an exponential potential, $V\left(
\phi\right)  =V_{0}e^{-\lambda\phi}$, and we find that equation admits as Lie
point symmetries the vector fields:
\[
X_{\pm}=a^{\pm\frac{\sqrt{6}}{2}\lambda-3}\exp\left(  \frac{\lambda\mp\sqrt
{6}}{2}\phi\right)  \left(  \pm\frac{\sqrt{6}}{6}a\partial_{a}+\partial_{\phi
}\right)
\]
and
\[
X_{0}=\frac{\lambda}{6}a\partial_{a}+\partial_{\phi}~,~X_{\Psi}=\Psi
\partial_{\psi}~
\]
in addition to the infinity number of symmetries which indicates the linearity
of the equation\footnote{For more details on the derivation and the
application of the Lie symmetries on the WdW equation see \cite{and1,and2} and
references cited therein.}. \ \ The vector fields, $X_{\pm},$ $X_{0}$, form
the Lie group $A_{3,3}$ which is commonly known as the semidirect product of
dilations and translations in the plane, i.e., $D\times_{s}T_{2}\equiv
A_{1}\times_{s}2A_{1}.$

We are interested on the vector field,~$X_{0}$, in which the normal
coordinates of $X_{0}$ are given from the transformation $a=\exp\left(
\frac{\lambda}{6}x\right)  ~,~\phi=x+y$. In the new variables the WdW equation
becomes%
\begin{equation}
\frac{6\Psi_{,xx}-12\Psi_{,xy}+\left(  6-\lambda^{2}\right)  \Psi_{,yy}%
}{2V_{0}\lambda^{2}}+e^{-\lambda y}\Psi=0.
\end{equation}
Consider now the vector field $Y_{1}=X_{1}+\mu_{1}\Psi\partial_{\psi}$, which
provides the invariant solution $\Psi\left(  x,y\right)  =\Phi\left(
y\right)  e^{\mu_{1}x},~$in which
\begin{equation}
\left(  6-\lambda^{2}\right)  \Phi_{,yy}-12\mu_{1}\Phi_{,y}+\left(  6\mu
_{1}^{2}+\frac{2V_{0}}{\lambda^{2}}e^{-\lambda y}\right)  \Phi=0.
\end{equation}
For $\lambda^{2}\neq6$ the solution of the last equation is given by%
\begin{equation}
\Phi\left(  y\right)  =e^{-\frac{\sqrt{6}}{2}C_{1}y}\left[  \Phi_{1}J_{C_{1}%
}\left(  Z\right)  +\Phi_{2}Y_{C_{1}}\left(  Z\right)  \right]  , \label{sol2}%
\end{equation}
where $C_{1}=-\frac{2\sqrt{6}\mu_{1}}{6-\lambda^{2}}$, $Z=\sqrt{\frac{8V_{0}%
}{6-\lambda^{2}}}e^{-\frac{\lambda}{2}y}$ and $J,Y$ are the Bessel functions
of the first and second kind respectively. We are not concerned with the
classical solution of the field equations. This can be found in \cite{Russo}.

Of course the general solution of the wavefunction is given from the sum of
all of the free parameters of solution (\ref{sol2}) and of the constant
$\mu_{1}$. However, in the following we consider the dominant term of the wavefunction.

Because a solution of the WdW equations is known we can derive the quantum
effects in the classical field equations as they are described in the
semiclassical approach of Bohmian Mechanics \cite{boh1} (for more details of
the method and for some recently applications see
\cite{kim,pinto,TchRN,manto,Stu2,sing1}). Specifically the quantum potential
is given from the formula $Q_{V}=-\frac{1}{2}\left(  \Delta A/A\right)  $, in
which $A$ is the amplitude of the wavefunction. The Bessel functions,
(\ref{sol2}), provide oscillatory terms in the wavefunction. Hence in order to
define the amplitude of the wavefunction we consider the behaviour of the
wavefunction $\Psi\left(  x,y\right)  $ at the limits.

In the following we assume that$~\operatorname{Re}\left(  \mu_{1}\right)  =0$,
and that $\lambda^{2}\neq6$, because in that case we have, $\ \Phi\left(
y\right)  =\exp\left(  \frac{3\mu_{1}^{2}y-\sqrt{6}V_{0}e^{-\sqrt{6}y}}%
{6\mu_{1}}\right)  $,~which means that the amplitude of the wavefunction is
constant, i.e., the quantum potential is zero.

For $\lambda^{2}\neq6$,~and $0<<Z<<C_{1}$, the wavefunction is approximated by
the function
\begin{equation}
\Psi\left(  x,y\right)  =\bar{\Phi}_{3}\cosh\left(  \left\vert \frac{C_{1}}%
{2}\lambda\right\vert y+y_{0}\right)  e^{\mu_{1}x}.
\end{equation}

From where we calculate that the quantum potential is constant,~$Q_{V}%
=\frac{\lambda^{2}-6}{8}\left(  C_{1}\right)  ^{2}=Q_{V}^{0}.~$The quantum
effects pass into the classical field equations and specifically into the
constraint equation (\ref{lang.02}) as follows~$\ \mathcal{H}+Q_{V}^{0}=0,$ or
equivalently,%
\begin{equation}
3\left(  \frac{\dot{a}}{a}\right)  ^{2}=\left(  \frac{1}{2}\dot{\phi}%
^{2}+V\left(  \phi\right)  \right)  +\frac{Q_{V}^{0}}{a^{3}}.
\end{equation}

The last term is nothing else than the component which corresponds to a dust
fluid term, which is minimally coupled to the scalar field, while the
continuous equation, $\dot{\rho}_{m}+3\frac{\dot{a}}{a}\rho_{m}=0,$ has been
solved. Recall that, as the quantum potential is constant, the second-order
field equations do not change. Here we can see that $Q_{V}^{0}=Q_{V}%
^{0}\left(  \lambda,\mu_{1}\right)  $, which means that the frequency of the
wave function is related to the energy density of the dust term.

In the other limit, in which $Z\rightarrow\infty$, the wavefunction is
approximated by
\begin{align}
e^{-\mu_{1}x}\Psi\left(  x,y\right)   &  =\bar{\Phi}_{1}\sqrt{\frac{1}{Z}}%
\cos\left(  Z-\frac{\left(  2C_{1}+1\right)  \pi}{4}\right)  +\nonumber\\
&  ~~~~+\bar{\Phi}_{2}\sqrt{\frac{1}{Z}}\sin\left(  Z-\frac{\left(
2C_{1}+1\right)  \pi}{4}\right)  .
\end{align}
Two cases have to be considered: $\lambda^{2}<6$, and $\lambda^{2}>6$.

For $\lambda^{2}<6$, the wavefunction becomes $\Psi\left(  x,y\right)
\simeq\frac{1}{\sqrt{Z}}\exp\left(  i\rho\left(  x,Z\right)  \right)  ~$from
which we derive the amplitude $A\left(  x,y\right)  =\frac{1}{\sqrt{Z}}$ and
we calculate the quantum potential to be constant as before. \ However, for
$\lambda^{2}>6$ things are different and we have that ~$\Psi\left(
x,y\right)  =e^{-\frac{\lambda}{4}y}e^{C_{2}e^{-\frac{\lambda}{2}y}}e^{\mu
_{1}x}$,~$C_{2}=\sqrt{\frac{8V_{0}}{\left\vert 6-\lambda^{2}\right\vert }.}%
$Therefore the quantum potential in the original coordinates is $\ $%
\begin{equation}
Q_{V}\left(  a,\phi\right)  =-\frac{1}{2}\left(  \frac{\lambda^{2}-32}%
{32}+V_{0}a^{3}e^{-\lambda\phi}\right)  ,
\end{equation}
where now except from the dust term which was introduced above there is also a
term which corresponds to the scalar-field potential.

The latter provides a dust term which satisfies the weak energy condition when
$\frac{\lambda^{2}}{32}<1$. However, in contrast to above the quantum
correction is independent of the frequency of the wavefunction. As far as
concerns the application of the invariant solutions which corresponds to the
vector fields $X_{\pm}$ in the semiclassical approximation, the quantum
potential which follows is zero. \

We studied the group invariant solutions of the WdW equation and we used them
to derive the quantum potential in the semiclassical approximation of Bohmian
mechanics. We show that the quantum potential provides a dust fluid component
in the scalar field cosmological model. Furthermore we found that the energy
density of the dust fluid depends upon the frequency of the wavefunction and
on the exponential power of the scalar-field potential. Moreover, the
introduction of the matter source excludes the singular solution of the
exponential potential \cite{Copeland1}.

For other kind of scalar field potentials that result does not necessary hold.
For instance for the hyperbolic potential which has been studied in
\cite{and1} from the solution of the WdW equation we find that the quantum
potential is zero. However the analysis above holds in the limit in which the
hyperbolic potential is approximated by the exponential potential.

Another important information that we can extract from the Lie group
invariants is the derivation of a family of boundary conditions in which they
are satisfied by the invariant solution. In particular in the boundary/initial
problem an invariant solution generated by a symmetry vector satisfies the
boundary/initial conditions iff the later are invariant under the action of
the same symmetry vector, see \cite{Bluman}.

\begin{acknowledgments}
This work was financial supported by FONDECYT grant no. 3160121. The author
thanks N. Dimakis for a fruitful discussion on the subject.
\end{acknowledgments}

\end{document}